\documentclass[journal=jpclcd,manuscript=letter, layout=twocolumn]{achemso}

\usepackage[version=5]{mhchem} 

\usepackage{natbib}
\usepackage{amsmath}
\usepackage{graphicx}

\author{Erik S. Thomson}
\email{erik.thomson@chem.gu.se}
\author{Xiangrui Kong}
\author{Patrik U. Andersson}
\affiliation[University of Gothenburg]
{Department of Chemistry, Atmospheric Science, University of Gothenburg, SE-412 96, Gothenburg, Sweden}

\author{Nikola Markovi\'{c}}
\affiliation[Chalmers University of Technology]
{Department of Chemical and Biological Engineering, Physical Chemistry, Chalmers University of Technology, SE-412 96, Gothenburg, Sweden}

\author{Jan B. C. Pettersson}
\email{janp@chem.gu.se}
\affiliation[University of Gothenburg]
{Department of Chemistry, Atmospheric Science, University of Gothenburg, SE-412 96, Gothenburg, Sweden}

\title {Collision Dynamics and Solvation of Water Molecules in a Liquid Methanol Film}

\begin{document}

\begin{abstract}
Environmental molecular beam experiments are used to examine water interactions with liquid methanol films at temperatures from 170 K to 190 K.  We find that water molecules with 0.32 eV incident kinetic energy are efficiently trapped by the liquid methanol.  The scattering process is characterized by an efficient loss of energy to surface modes with a minor component of the incident beam that is inelastically scattered.  Thermal desorption of water molecules has a well characterized Arrhenius form with an activation energy of 0.47$\pm 0.11$ eV and pre-exponential factor of $4.6 \times 10^{15\pm3} $ s$^{-1}$.  We also observe a temperature dependent incorporation of incident water into the methanol layer.  The implication for fundamental studies and environmental applications is that even an alcohol as simple as methanol can exhibit complex and temperature dependent surfactant behavior.  
\end{abstract}

\begin{figure}
\centering
\includegraphics*[width=1\columnwidth, clip=true]{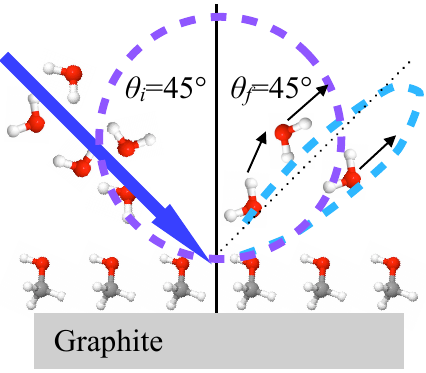}
\label{fig:skem}
\end{figure}
\begin{center}
\small{\textbf{Keywords:} Methanol, Molecular Beam, Water Uptake, Desorption Kinetics,\\
 Collision Dynamics, H/D exchange}
\end{center}
\normalsize

Methanol (CH$_3$OH) differs from water only due to the interchange of a methyl group for a single hydrogen.  However, the effect is strong and can be particularly interesting when the two compounds interact.  Previous studies of CH$_3$OH - H$_2$O interactions have focused on liquids \cite{Jayne1991}, ices \cite{Wolff2007}, amorphous solid water \cite{Bahr2008}, and a range of environments.  Particular interest has focused on uptake and sticking coefficients for methanol on ice and liquid surfaces \cite{Jayne1991, Morita2003} and its ensuing surfactant effects \cite{Hudson2002, Bahr2008}.  Molecular dynamics simulations show that methanol molecules strongly interact with H$_2$O surfaces and that the methanol-water interaction can be stronger than the methanol-methanol attraction \cite{Picaud2000}.  One way to think about the methanol-water interaction is as a competition between the hydroxyl group's affinity for hydrogen bonding and the hydrophobic nature of the methyl group \cite{Morita2003}.  These different interactions can lead to stable organic monolayers on ice and water \cite{Picaud2000}.  

In the atmosphere, the interaction of gas-phase molecules and surfaces has wide ranging effects for physical and chemical process like cloud formation and photo-chemistry.  Alcohol coated surfaces may be important and substantial sources or sinks for HO$_x$ radicals, especially in the dry upper troposphere where the lack of water vapor limits its production through ozone photolysis \cite{Winkler2002, Hudson2002}.  Atmospheric methanol is a surfactant of particular interest because of its ubiquity \cite{Hudson2002}, and because it also serves as a simple model for longer aliphatic molecules.  Alcohol coverages may limit atmospheric particle growth because they may render surfaces somewhat hydrophobic.  However, this is still controversial, as mass accommodation coefficients and effects of surfactant properties on mass transfer through surface layers remain poorly constrained.  Experimental measurements of size selected water droplets interacting with CH$_3$OH vapor have found mass accommodation coefficients as low as $\approx0.06$ \cite{Jayne1991} while dynamical simulations of CH$_3$OH molecules impinging on pure water surfaces suggest coefficients of order unity \cite{Morita2003}.  

The importance of methanol-water interactions has motivated our fundamental experimental studies of the molecular level dynamics of such systems.  Here we report findings from environmental molecular beam (EMB) experiments where supersonic D$_2$O molecules collide with a thin liquid-like methanol layer with temperatures $T_s$ between 170 K and 190 K.  The EMB apparatus, whose design allows us to probe surfaces under higher, environmentally relevant, vapor pressures than are accessible with standard MB technology, has been described in detail previously \cite{Andersson2000a, Suter2006, Kong2011}.  It consists of differentially pumped vacuum chambers to achieve a central ultra-high vacuum (UHV).  Methanol condenses on a graphite surface that is immediately surrounded by an inner chamber enabling a small region of finite vapor pressure and allowing for the formation of stable methanol surface films in dynamical equilibrium with their vapor, while simultaneously minimizing the molecular beam's transmission attenuation.  In contrast with traditional molecular beam experiments that are preformed under strict UHV conditions, the inner chamber allows us to maintain methanol vapor pressures in the $10^{-3}$ mbar range.  The D$_2$O is added to a He gas beam to increase its kinetic energy and allow us to simultaneously probe methanol surface coverage.  The incident kinetic energy (0.32 eV) results in measurable inelastic scattering, allowing us to probe collision dynamics, while monitoring He elastic scattering from the graphite substrate ensures complete methanol coverage throughout experiments \cite{Kong2011}.  Pulses of the gas beam are synchronized with a frequency chopper to select the central portion of each pulse, producing discrete 400 $\mu$s  gas impulses.  Within the UHV chamber a differentially pumped quadrupole mass spectrometer (QMS) ionizes particles leaving the surface by electron bombardment.  Detected time versus ion intensity counts are processed and output by a multi-channel scaler with a dwell time of 10 $\mu$s.   With the known experimental geometry the measured arrival intensities are easily translated into time-of-flight (TOF) measurements for particles traveling, within the plane defined by the beam and surface normal, from the surface to the detector.  Thus the TOF distributions can be analyzed to illuminate the important surface processes \cite{Scoles1988}.   

Within our experimental temperature range, previous X-ray diffraction studies have shown thin layers of methanol to be liquid \cite{Morishige1990}.  The melting point for a single monolayer is 135 K and increases with increasing film thickness to the bulk melting temperature of 175.4 K.  Monolayer methanol coverage has also been shown to have higher desorption energies than subsequent layers \cite{Bolina2005a, Ulbricht2006}, implying that the first layer of CH$_3$OH completely wets the graphite and adheres strongly.  Here we adjust the pressure to be high enough to maintain a complete CH$_3$OH layer on the graphite surface that we monitor by measuring elastic He scattering\cite{Scoles1992}, and maintain a low enough pressure to avoid multi-layers, observed with a light reflection technique.  Film thickness can be continuously monitored by observing interference from the reflections of a 670 nm laser.  In these experiments no beam attenuation and therefore no methanol film growth was observed, thus ensuring that for the experimental temperatures the CH$_3$OH remains a thin layer.  Heavy water is substituted for H$_2$O to enhance the signal-to-noise ratio and highly oriented pyrolythic graphite (HOPG, grade ZYB) is used as a substrate.  The use of HOPG is beneficial due to its well characterized helium scattering properties \cite{Scoles1992}, its well studied interactions with methanol layers \cite{Morishige1990, Bolina2005a, Wolff2007}, and its utility as an analog for atmospheric particles like black carbon \cite{Perraudin2007}.  

\begin{figure}
\centering
\includegraphics[width=1\columnwidth, clip=true]{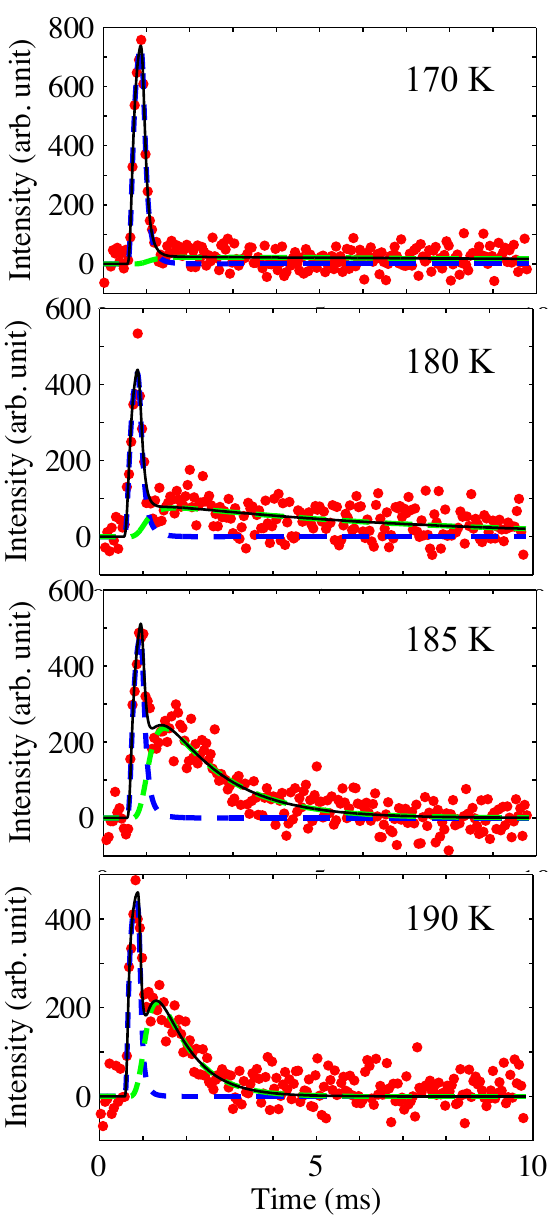}
\caption{Measured TOF distributions from D$_2$O incident on a methanol covered graphite surface.  Red points are a five point step-wise average of experimental data.  The solid black curve represents the fitted distribution with inelastic and trapping-desorption components represented by the blue and green dashed curves, respectively. }
\label{fig:dists}
\end{figure}

We have systematically characterized D$_2$O interactions with liquid methanol layers and compared with the bare graphite surface.  For all experiments the incident beam angle and measured angle of reflection are limited to $45^\circ$, due to the constraints of the inner-most chamber.  Methanol pressures in the inner chamber directly above the 185 K surface are estimated to be $\approx 2\times10^{-3}$ mbar from monitoring the UHV chamber pressure.  For surface temperatures below 185 K the added methanol does not significantly contribute to the UHV background making an inner chamber pressure estimate impossible.  \ref{fig:dists} compares the TOF distributions of D$_2$O scattered from methanol covered graphite for different surface temperatures.  
A fitted distribution is plotted above the recorded data, in addition to the individual fits for the inelastic and trapping-desorption components.  Clearly, D$_2$O collisions with the methanol covered surface exhibit inelastic scattering and trapping-desorption behaviors.  At low temperatures the inelastic component of the scattered intensity dominates the signal.  Above 180 K this changes significantly and thermally activated molecules more quickly desorb from the surface.

The non-linear least squares fitting and quantitative analysis of the final TOF distributions can be summarized as a convolution of the initial beam distribution with a component of inelastically scattered particles and another component of thermally desorbed particles.  The initial beam distribution is measured directly by rotating the QMS into the beam path.  Theoretical inelastic and trapping-desorption distributions are calculated and separately convoluted with the incident beam.  Finally, using a non-linear fitting algorithm a linear combination of these distributions is used to theoretically fit the measured data.  For these experiments we assume first-order thermal desorption with a residence time behavior of the form, 
\begin{equation}
\label{eq:desorp} F_{res}(t)={C}_1\exp(-kt).
\end{equation}
Here ${C}_1$ is a fitted scaling factor, $t$ is the surface residence time, and $k$ is the desorption rate constant.  The inelastic scattering distribution is also assumed to have the common form \cite{Suter2006}, 
\begin{equation}
\label{eq:Iis} I_{is}(t)={C}_2 v^4 \exp\left[{-\left(\frac{v-\bar{v}}{v_{is}}\right)^2}\right],
\end{equation}
where ${C}_2$ is a second fit scaling factor, $v$ is the particle velocity calculated from the measurement time and flight path length,  $\bar{v}$ is the average inelastically scattered beam velocity, and $v_{is}$ is, 
\begin{equation}
\label{eq:vbeam} v_{is}=\sqrt{\frac{ 2k_B T_b}{m}}.
\end{equation}
The temperature $T_b$ represents the inelastically scattered beam temperature, $k_B$ is the Boltzmann constant, and $m$ is the molecular mass in kilograms.  Both $\bar{v}$ and $T_b$ are left as free fitting parameters when assuming an inelastic contribution.  At the lowest experimental surface temperatures $T_s \leq 170$ K trapping-desorption occurs on long time scales merging with the background.  Thus the recorded signal is primarily due to the inelastically scattered component.  As the temperature increases D$_2$O more efficiently desorbs from the surface, shrinking the exponential's tail but increasing the contribution of trapping-desorption to the measured signal (\ref{fig:dists}). 

\begin{figure}
\centering
\includegraphics[width=1.0\columnwidth, clip=true]{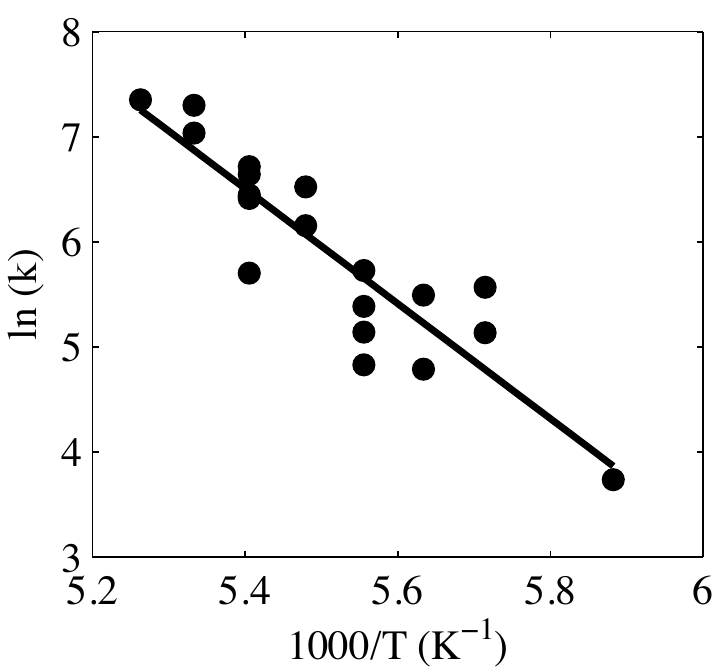}
\caption{Arrhenius plot of the rate coefficients for desorption of D$_2$O from liquid methanol.  The solid line is a linear least-squares fit to the points with a slope of $E_A=0.47\pm0.11$ eV resulting in a pre-exponential factor $A=4.6 \times 10^{15\pm3}$ s$^{-1} $.}
\label{fig:Arh}
\end{figure}

The temperature dependence of the desorption rate coefficient is summarized in \ref{fig:Arh}.  
The linear response of \ref{fig:Arh} demonstrates that the desorption kinetics of D$_2$O from methanol do exhibit Arrhenius type behavior, $k=A \exp (-E_A/k_BT_s)$.  The resulting activation energy $E_A=0.47\pm0.11$ eV and pre-exponential factor $A=4.6 \times 10^{15\pm3} $ s$^{-1}$ with their respective 95\% confidence intervals, are in good agreement with previous measurements of kinetic parameters for thin layers of pure H$_2$O and CH$_3$OH \cite{Ulbricht2006}.   This result is not unanticipated because the hydrogen bonds associated with both methanol and water are expected to place the dominant constraint on their surface behavior.  Various transition state theory models have predicted that such adsorbate interactions result in comparable pre-exponential factors \cite{Ulbricht2006,Seebauer1988}.  Thus the desorption behavior of water from methanol is similar to the desorption of either compound from itself.   

\begin{figure}
\centering 
\includegraphics[width=1.0\columnwidth, clip=true]{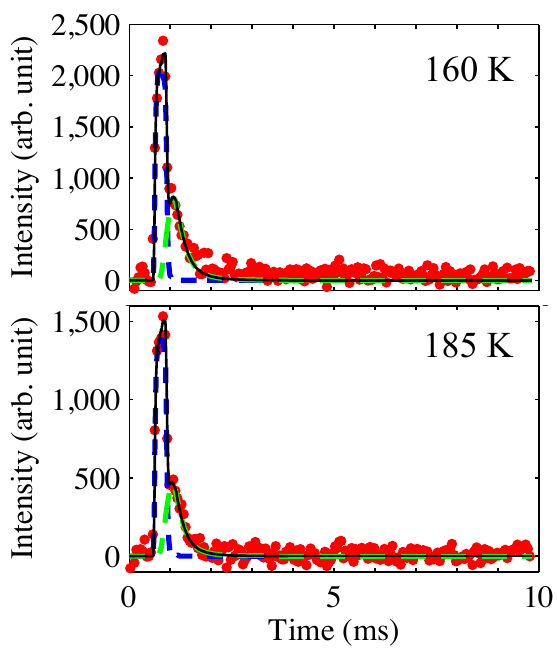}
\caption{Measured TOF distributions (c.f., \ref{fig:dists}) for D$_2$O incident on bare graphite.}
\label{fig:baregr}
\end{figure}

In contrast to the liquid methanol, thermal desorption of D$_2$O from bare graphite is very fast.  \ref{fig:baregr} shows that for bare graphite desorption curves do not vary with temperature and $k>>10^3$ s$^{-1}$.
Directly comparing the inelastic scattering components for the graphite and methanol surfaces is difficult for a single scattering angle, due to the fact that the angular distributions depend upon the type of surface.  However, comparison with earlier studies of water interactions with graphite\cite{Markovic1999} suggests that the inelastic component is small relative to trapping-desorption.  The average final kinetic energy of scattered molecules was 10\% of the incident energy for the methanol-covered surface and 55\% for the bare graphite, independent of surface temperature.  The results for bare graphite are in good agreement with the results for the H$_2$O-graphite system \cite{Markovic1999}.  For the methanol film the results confirm that D$_2$O collisions are highly inelastic and characterized by very efficient transfer of energy to surface modes.  For comparison, 20-25\% of the kinetic energy is conserved by Ar, HCl, and H$_2$O molecules with similar incident parameters, scattering from water ice \cite{Andersson2000a, Andersson2000b, Gibson2011}.

\begin{figure}
\centering 
\includegraphics[width=1.0\columnwidth, clip=true]{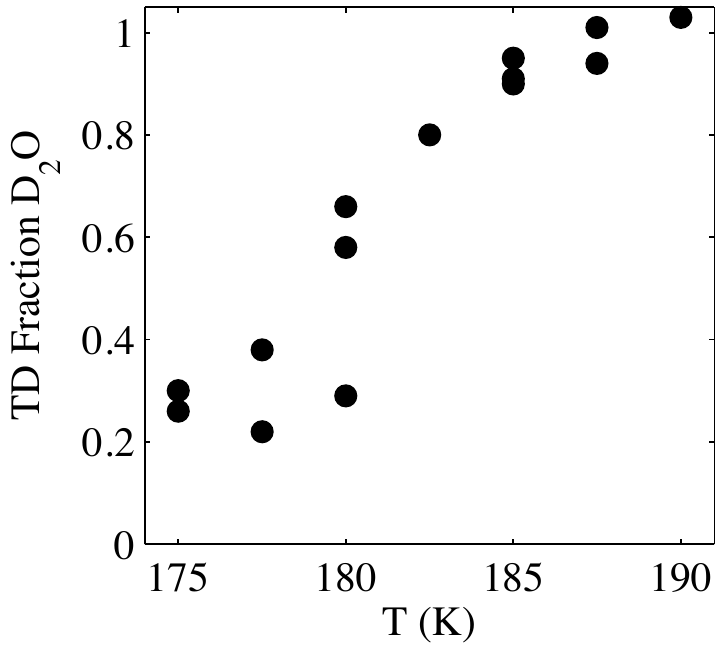}
\caption{The fraction of trapping-desorbed (TD) D$_2$O molecules relative to the incident number as a function of temperature.  For clarity error bars are omitted and an explanation of the uncertainty is restricted to the main text.}
\label{fig:TDfrac}
\end{figure}

With the help of detailed classical molecular dynamic simulations, the trapping-desorption distributions from bare graphite can be used to estimate the trapping efficiency of the methanol covered surface.  We performed new calculations focusing on the trapping probability of D$_2$O incident on a bare graphite surface in an identical manner to previously published results for H$_2$O \cite{Markovic1999}.  Molecules incident at $45^\circ$ on a 180 K surface with kinetic energies of 0.32 eV were simulated to have an 80\% chance of being trapped.  Using this result we calculated an incident beam intensity from the trapping-desorption of the bare graphite case.  Normalizing for beam attenuation due to higher vapor pressures in the methanol experiments we computed the fraction of incident molecules measured in the liquid methanol trapping-desorption distributions and plot them in \ref{fig:TDfrac}.  An uncertainty of $\approx \pm 20 \%$ in the absolute values plotted in \ref{fig:TDfrac} results from the limitations of the simulated trapping and experimental measurements.  However, such uncertainties are systematic and therefore the strong observed trend with temperature persists without regard to the absolute ratio.

\ref{fig:TDfrac} shows that there is a clear trend in the trapping-desorption fraction as a function of temperature.  At high temperatures almost all incident molecules are trapped and subsequently desorbed from the methanol surface.  At low temperatures as few as 20\% of the molecules are thermally desorbed from the surface, leaving a large unaccounted for water reservoir and suggesting that on the time scale of the experiment (10 ms) some D$_2$O is lost within the methanol layer.  

In this case two possible D$_2$O sinks exist.  First, there may exist some level of H/D isotopic exchange between the methanol layer and the D$_2$O beam.  Previously rapid H/D exchange has been measured for cryogenic methanol systems at temperatures above 150 K \cite{Souda2003}.  For CH$_3$OH interacting with a D$_2$O ice layer \citet{Souda2003} found almost complete H/D exchange.  However, for the reverse case of D$_2$O adsorbed on methanol surfaces H/D exchange was not explicitly observed.  Rather above 120 K their results suggested that D$_2$O either dissolved into the bulk methanol by forming hydrogen bonds, or formed islands that were subsequently covered by CH$_3$OH.  In our experiments formation and desorption of HDO could serve as an indicator of H/D exchange taking place within the methanol layer.  However, we were unable to observe H/D exchange over the entire temperature range.  Thus the noise of the experimental measurements limited the maximum HDO formation to less than 1\% of the incident molecules.  This observation is further supported by measurements of H/D exchange on mineral \cite{Hsieh1999} and liquid \cite{Dempsey2011} surfaces and for H$_2$O/CD$_3$OD mixtures at up to 170 K \cite{Ratajczak2009} that indicate time scales of minutes to hours and longer for significant isotopic exchange.  It is likely that desorption of isotopically light water in a CH$_3$OH-D$_2$O system would be thermally activated at long time scales, similar to what is observed for acids and cold salty water solutions \cite{Brastad2011}.  We conclude that below 185 K water, which only desorbs on long timescales, is incorporated into the methanol layer.  Such a process would only contribute to the background D$_2$O levels of the measurements, and thus be negated during subsequent analysis.  

We have studied water interactions with liquid methanol layers on a graphite substrate between 170 K and 190 K.  At these temperatures, the methanol surface layer was maintained in a dynamic state with the help of a finite vapor pressure above the surface.  Collisions between water molecules and the liquid methanol layer were observed to result in efficient surface trapping with only a small fraction of the hyperthermal incident molecules inelastically scattered.  The escaping molecules had lost more than 80\% of their incident kinetic energy, indicating a very efficient energy transfer to surface modes.  The desorption kinetics have an Arrhenius type behavior and the activation energy we have calculated is indicative of multiple hydrogen bonds between D$_2$O and CH$_3$OH molecules within the liquid \cite{Beta2005}.  On the millisecond time scale of the experiments desorption competes with loss of D$_2$O to more strongly bound states within the layer, and high temperature is observed to favor desorption.  Loss of D$_2$O due to H/D exchange is likely less important since no desorbing HDO was detected in the experiments.  
 
This study contributes to the fundamental understanding of gas accommodation and uptake in organic liquids and is of potential importance for the description of the effect of organic surfactants on heterogeneous processes in the atmosphere and in other environments.  One immediate implication is that the effect of CH$_3$OH as a common atmospheric surfactant will be temperature dependent and may even contribute to water uptake by otherwise hydrophobic particles.  This provides context for continued studies of more complicated surfactants of environmental importance, such as longer chain alcohols.

\begin{acknowledgement}

Funding for this research was provided by the Swedish Research Council and the University of Gothenburg.  We thank the anonymous referees whose suggestions improved this letter.

\end{acknowledgement}

\providecommand*\mcitethebibliography{\thebibliography}
\csname @ifundefined\endcsname{endmcitethebibliography}
  {\let\endmcitethebibliography\endthebibliography}{}

\end{document}